\documentclass[letterpaper,twocolumn,prl,aps,superscriptaddress,floatfix]{revtex4}

\usepackage[latin9]{inputenc}
\usepackage{amsmath}
\usepackage{amssymb}
\usepackage{graphicx}
\usepackage{esint}
\usepackage{times}
\usepackage{ulem}
\usepackage{pifont}

\usepackage[unicode=true,
bookmarks=true,bookmarksnumbered=false,bookmarksopen=false,
breaklinks=false,pdfborder={0 0 1},backref=false,colorlinks=true]
{hyperref}
\hypersetup{
    linkcolor=magenta, urlcolor=blue, citecolor=blue, pdfstartview={FitH}, hyperfootnotes=false, unicode=true}

\makeatletter

\pdfpageheight\paperheight
\pdfpagewidth\paperwidth

\@ifundefined{textcolor}{}
{%
    \definecolor{BLACK}{gray}{0}
    \definecolor{WHITE}{gray}{1}
    \definecolor{RED}{rgb}{1,0,0}
    \definecolor{GREEN}{rgb}{0,1,0}
    \definecolor{BLUE}{rgb}{0,0,1}
    \definecolor{CYAN}{cmyk}{1,0,0,0}
    \definecolor{MAGENTA}{cmyk}{0,1,0,0}
    \definecolor{YELLOW}{cmyk}{0,0,1,0}
}

\usepackage{xcolor}
\usepackage{soul}

\definecolor{blue}{rgb}{0,0,1}
\definecolor{red}{rgb}{1,0,0}
\definecolor{green}{rgb}{0,1,0}

\makeatother

\begin{document}

\title{Autonomous stabilization of Fock states in an oscillator against multiphoton losses}

\author{Sai Li}
\thanks{These authors contributed equally to this work.}
\affiliation{Shenzhen Institute for Quantum Science and Engineering, Southern University of Science and Technology, Shenzhen 518055, China}
\affiliation{International Quantum Academy, Shenzhen 518048, China}
\affiliation{Guangdong Provincial Key Laboratory of Quantum Science and Engineering, Southern University of Science and Technology, Shenzhen 518055, China}

\author{Zhongchu Ni}
\thanks{These authors contributed equally to this work.}
\affiliation{Shenzhen Institute for Quantum Science and Engineering, Southern University of Science and Technology, Shenzhen 518055, China}
\affiliation{International Quantum Academy, Shenzhen 518048, China}
\affiliation{Guangdong Provincial Key Laboratory of Quantum Science and Engineering, Southern University of Science and Technology, Shenzhen 518055, China}
\affiliation{Department of Physics, Southern University of Science and Technology, Shenzhen 518055, China}

\author{Libo Zhang}
\author{Yanyan Cai}
\author{Jiasheng Mai}
\author{Shengcheng Wen}

\author{Pan Zheng}
\author{Xiaowei Deng}
\affiliation{Shenzhen Institute for Quantum Science and Engineering, Southern University of Science and Technology, Shenzhen 518055, China}
\affiliation{International Quantum Academy, Shenzhen 518048, China}
\affiliation{Guangdong Provincial Key Laboratory of Quantum Science and Engineering, Southern University of Science and Technology, Shenzhen 518055, China}

\author{Song Liu}
\email{lius3@sustech.edu.cn}

\author{Yuan Xu}
\email{xuyuan@iqasz.cn}
\affiliation{Shenzhen Institute for Quantum Science and Engineering, Southern University of Science and Technology, Shenzhen 518055, China}
\affiliation{International Quantum Academy, Shenzhen 518048, China}
\affiliation{Guangdong Provincial Key Laboratory of Quantum Science and Engineering, Southern University of Science and Technology, Shenzhen 518055, China}
\affiliation{Shenzhen Branch, Hefei National Laboratory, Shenzhen 518048, China}

\author{Dapeng Yu}
\affiliation{Shenzhen Institute for Quantum Science and Engineering, Southern University of Science and Technology, Shenzhen 518055, China}
\affiliation{International Quantum Academy, Shenzhen 518048, China}
\affiliation{Guangdong Provincial Key Laboratory of Quantum Science and Engineering, Southern University of Science and Technology, Shenzhen 518055, China}
\affiliation{Department of Physics, Southern University of Science and Technology, Shenzhen 518055, China}
\affiliation{Shenzhen Branch, Hefei National Laboratory, Shenzhen 518048, China}

\begin{abstract}
Fock states with a well-defined number of photons in an oscillator have shown a wide range of applications in quantum information science. Nonetheless, their usefulness has been marred by single and multiple photon losses due to unavoidable environment-induced dissipation. Though several dissipation engineering methods have been developed to counteract the leading single-photon loss error, averting multiple photon losses remains elusive. Here, we experimentally demonstrate a dissipation engineering method that autonomously stabilizes multi-photon Fock states against losses of multiple photons using a cascaded selective photon-addition operation in a superconducting quantum circuit. Through measuring the photon-number populations and Wigner tomography of the oscillator states, we observe a prolonged preservation of nonclassical Wigner negativities for the stabilized Fock states $\vert N\rangle$ with $N=1,2,3$ for a duration of about $10$~ms. Furthermore, the dissipation engineering method demonstrated here also facilitates the implementation of a non-unitary operation for resetting a binomially-encoded logical qubit. These results highlight potential applications in error-correctable quantum information processing against multi-photon-loss errors. 
\end{abstract}

\maketitle
\vskip 0.5cm
Bosonic modes, such as harmonic oscillators, have garnered significant attention in various fields including quantum optics, quantum metrology, quantum simulation, and quantum computation~\cite{gu2017, terhal2020, cai2021,joshi2021, ma2021, krasnok2023}. 
These applications require precise and robust generation of nonclassical bosonic states in the oscillators,
of which number states or Fock states are the most fundamental. Fock states are energy eigenstates with a fixed number of excitations or bosons, which possess a pivotal role in quantum optics, quantum metrology, and quantum information science~\cite{Aaron2020}. 
Over the past few decades, significant efforts have been made to create and manipulate Fock states in various physical platforms, including motional modes of trapped ions~\cite{meekhof1996}, mechanical oscillators~\cite{arrangoiz2019, wollack2022}, microwave cavities with Rydberg atoms~\cite{sayrin2011}, acoustic wave resonators~\cite{chu2018, satzinger2018, vonlupke2022}, and superconducting circuits~\cite{houck2007, schuster2007, hofheinz2008, wang2011, heeres2015, premaratne2017, wang2017, kudra2022}. 
However, all these methods still manifest significant challenges, largely stemming from multiple energy quanta losses (such as multi-photon losses) in the oscillator arising from non-negligible coupling to the uncontrolled dissipative environment.

Dissipation, on the other hand, can be well controlled and is an essential resource for quantum information processing~\cite{harrington2022}. Subtly engineered dissipation has found wide-ranging applications in preparing and stabilizing quantum states~\cite{murch2012, kienzler2015, holland2015, leghtas2015, lu2017, touzard2018}, resetting quantum system~\cite{geerlings2013, bultink2016, magnard2018, egger2018, zhou2021}, generating entanglement~\cite{krauter2011,shankar2013, kimchi2016, liu2016, andersen2019}, implementing quantum simulation~\cite{ma2019}, and performing quantum error correction~\cite{ofek2016, gertler2021, ni2023, sivak2023}. In these examples, dissipation generally serves as a one-way valve to remove the system entropy to a reservoir bath and irreversibly transfers the system into a dark state or dissipation-free manifold. 
In the context of superconducting circuit quantum electrodynamics (QED) systems~\cite{blais2021}, a quantum reservoir-engineering method has been employed to implement autonomous stabilization of a single-photon Fock state in a microwave cavity~\cite{holland2015}. Nevertheless, extending this method to stabilize higher-photon-number states against multi-photon losses remains a formidable experimental challenge.
Recently, engineered dissipation has been applied to autonomous quantum error correction through a parity recovery operation for stabilizing photon number parity against single-photon-loss error~\cite{gertler2021}. 
However, dissipation-induced multi-photon losses continue to persist as a major source of error that jeopardizes the integrity of quantum states in the oscillator and has yet to be fully resolved.

In this Letter, we experimentally demonstrate a dissipation engineering method for autonomously stabilizing multi-photon Fock states against multi-photon-loss errors in a superconducting microwave cavity. The robust generation of these nonclassical bosonic states is realized by developing a cascaded selective photon-addition (CSPA) operation assisted by an ancillary superconducting qubit and a reservoir bath resonator. The non-unitary CSPA operation is implemented by applying a continuous-wave multi-component microwave drive to engineer the coupling to the dissipative environment. 
The experimental results indicate that the nonclassical Wigner negativities of the Fock states are preserved for a duration of about $10$~ms, which can potentially be extended indefinitely. By controlling the selective photon-addition operations on individual photonic energy levels, we have confirmed that multi-photon-loss errors have been corrected to stabilize multi-photon Fock states. Furthermore, we showcase the potential of our dissipation engineering method in implementing a non-unitary quantum operation that resets a binomially-encoded logical qubit.

\begin{figure}[tb]
    \includegraphics{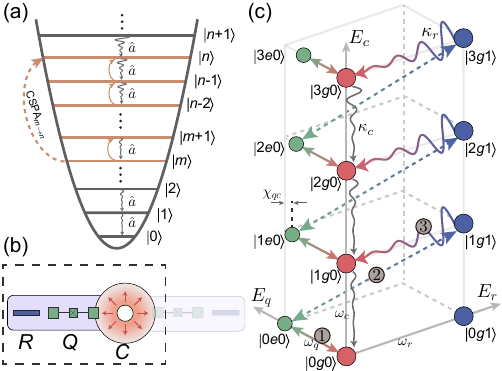}
    \caption{Schematic of the multi-photon Fock state stabilization.
 (a) Fock state $|n\rangle$ in a harmonic oscillator is stabilized against losses of $n-m$ photons by continuously implementing the non-unitary operation $\mathrm{CSPA}_{m\rightarrow n}$ in a truncated photon number subspace. The CSPA operation comprises a series of selective photon addition operations between adjacent energy levels. 
(b) Experimental device schematic consisting a 3D microwave cavity $C$ as the oscillator, which is dispersively coupled to two superconducting qubits $Q$ and associated readout resonators $R$. Note that only the left qubit and its readout resonator (inside dashed box) are used in the experiment.
 (c) Energy level diagram of the $C$-$Q$-$R$ system. The CSPA operation is realized by combining two sets of continuous-wave frequency combs and the fast spontaneous decay of the readout resonator. The first set of comb selectively drives the qubit transitions (\ding{172}) and the second set swaps the excitation of the qubit to the storage cavity and readout resonator (\ding{173}). The fast spontaneous decay dissipates the excitation of $R$ to the environment (\ding{174}), thus stabilizing the Fock states in cavity $C$. }         
    \label{fig1}
\end{figure}

As shown in Fig.~\ref{fig1}(a), a quantum harmonic oscillator with a quadratic potential well has infinite evenly-spaced energy levels. Each energy level corresponds to an energy eigenstate $|n\rangle $, commonly known as the photon number state or Fock state, containing $n$ photons in the oscillator. However, these Fock states are susceptible to environment-induced decoherence, resulting in photon loss described by the annihilation operator $\hat{a} = \sum_n{\sqrt{n} |n-1\rangle \langle n |}$. To stabilize the Fock state against decoherence-induced photon loss errors, we introduce a CSPA operation, capable of adding photons in a tailored photon number subspace. This operation is similar to the conjugate operator of $\hat{a}$ and can be described by
\begin{eqnarray} 
\mathrm{CSPA}_{m\rightarrow n} = \sum_{i=m}^{n-1}{|i+1\rangle \langle i |}
\end{eqnarray}
in a truncated photon number subspace. Continuously implementing CSPA operations adds $n-m$ photons to an initial Fock state $| m \rangle$, resulting in the target Fock state $| n \rangle$. This non-unitary operation is implemented through simultaneously applying selective photon-addition operators $\hat{\Lambda}_{i\rightarrow i+1} = |i+1\rangle \langle i |$ between adjacent energy levels, adding a single photon to Fock state $|i\rangle$. These cascaded operations finally stabilize the Fock state $|n\rangle$ against losses of $n-m$ photons.

In our experiment, the non-unitary CSPA operation is implemented in a superconducting quantum circuit system, similar to that in Ref.~\cite{ni2022}, with the schematic shown in Fig.~\ref{fig1}(b). The system comprises a three-dimensional microwave cavity ($C$), acting as the oscillator, and two superconducting transmon qubits ($Q$) and their associated readout resonators ($R$). Note that only one qubit and its associated readout resonator are utilized to implement the quantum reservoir control for stabilizing multi-photon Fock states in cavity $C$. The relevant leading-order interactions of the $C$-$Q$-$R$ system can be described by the dispersive Hamiltonian:
\begin{eqnarray} 
\label{H0}
H_0/\hbar & = & \omega_c \hat{a}^\dagger \hat{a} + \omega_q |e\rangle \langle e | + \omega_r \hat{r}^\dagger \hat{r} \notag\\
        & - & \chi_{qc} \hat{a}^\dagger \hat{a} |e\rangle \langle e | - \chi_{qr} \hat{r}^\dagger \hat{r} |e\rangle \langle e |,
\end{eqnarray}
where $\hat{a}$ and $\hat{r}$ are the annihilation operators of the cavity and readout resonator modes, respectively. $| e\rangle$ and $| g \rangle$ represent the excited and ground states of the transmon qubit, respectively. The resonance frequencies of the three modes are $\omega_c/2\pi = 6.37$~GHz, $\omega_q/2\pi=5.68$~GHz, and $\omega_r/2\pi=8.60$~GHz. The dispersive couplings between the qubit and the other two modes are $\chi_{qc}/2\pi=7.7$~MHz and $\chi_{qr}/2\pi=2.4$~MHz. The microwave cavity has a single-photon lifetime of about 50~$\mu$s (corresponding to a decay rate $\kappa_c/2\pi = 3.1$~kHz) for storing photons, while the readout resonator is designed with a fast decay rate of $\kappa_r/2\pi = 2.4$~MHz for qubit readout and assisting in dissipation engineering. More details of the device parameters can be found in the Supplementary Material~\cite{supplement}.

With the relevant energy levels of the system illustrated in Fig.~\ref{fig1}(c), the CSPA operation is achieved by applying two sets of continuous-wave drives, resulting in a drive Hamiltonian:
\begin{eqnarray} 
\label{Hd}
H_d/\hbar &=& \sum_{i=m}^{n-1}{\left(\Omega_i |i,e,0\rangle \langle i,g,0| + J_i |i+1,g,1\rangle \langle i,e,0| \right)} \notag\\
&+& \mathrm{H.c.},
\end{eqnarray}
under the rotating wave approximation. Here the indices in the bras and kets are labeled by the Fock states in the $C$-$Q$-$R$ order, and $\mathrm{H.c.}$ represents Hermitian conjugate. By exploiting the photon-number-dependent qubit frequency shift in the dispersive Hamiltonian described in Eq.~(\ref{H0}), we can achieve selective transitions between individual energy levels.

\begin{figure}[tb]
    \includegraphics{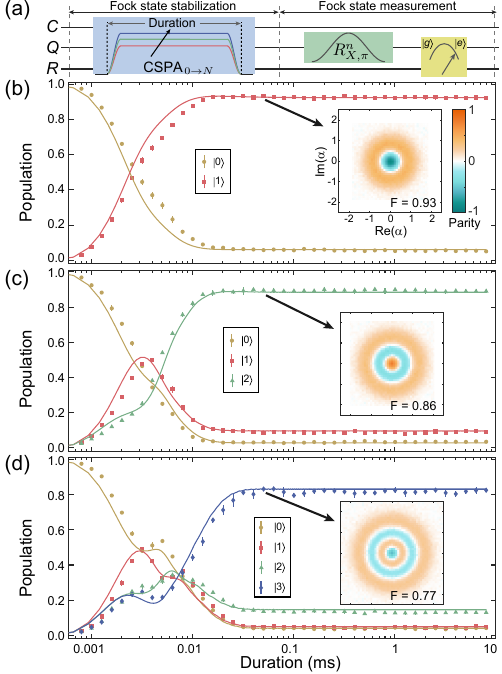}
    \caption{Autonomous stabilization and characterization of multi-photon Fock states.
	(a) Experimental sequence for measuring the photon number populations of the stabilized Fock states.
   (b)-(d) Measured photon number populations (symbols) as a function of the evolution duration of the non-unitary CSPA operation, for stabilizing the Fock states $|1\rangle$ (b),  $|2\rangle$ (c), and $|3\rangle$ (d), respectively. Error bars are standard deviations from repeated experiments. Solid lines are from numerical simulations. The insets in (b-d) show the Wigner snapshots of the stabilized Fock states at a duration of about 50 $\mu$s.}
    \label{fig2}
\end{figure}

The drive Hamiltonian described in Eq.~(\ref{Hd}) comprises two sets of continuous-wave frequency combs, as illustrated with double arrows in the energy level diagram in Fig.~\ref{fig1}(c). The first set of drives achieves resonant oscillations between the states $|i,g,0\rangle$ and $|i,e,0\rangle$ at a coupling rate of $\Omega_i$. The second set targets the transition $|i,e,0\rangle \leftrightarrow |i+1,g,1\rangle$ at a coupling rate of $J_i$ by applying a four-wave mixing pump on the superconducting qubit~\cite{supplement}. Finally, the fast spontaneous decay of the reservoir resonator $R$ effectively removes the entropy by irreversibly transferring the state $|i+1,g,1\rangle$ to $|i+1,g,0\rangle$ with a loss rate of $\kappa_r$. As a result, combined with the two sets of drives, a photon-addition operator $\hat{\Lambda}_{i\rightarrow i+1} $ is constructed for adding a single photon to the Fock state $|i\rangle$ in $C$, given that $\kappa_c < \Omega_i < J_i < \kappa_r <\chi_{qc}$. By selectively applying a series of these drives, we can achieve the non-unitary operation $\mathrm{CSPA}_{m\rightarrow n}$ to autonomously stabilize the Fock state $|n\rangle$ against the loss of $n-m$ photons in the cavity.  

In our experiment, we first demonstrate the non-unitary CSPA operations to autonomously generate Fock states $|N\rangle$ in the storage cavity. The experimental sequence is depicted in Fig.~\ref{fig2}(a), where we first apply a $\mathrm{CSPA}_{0\rightarrow N}$ operation on an initial vacuum state in the cavity for a variable duration, and then perform a transmon spectroscopy measurement to extract the photon number populations in the cavity after the CSPA operation. 

Figures~\ref{fig2}(b)-\ref{fig2}(d) show the measured photon number populations as a function of the total evolution duration of the CSPA operation for generating Fock states $|N\rangle$ with $N=1,2,3$. The insets show the measured Wigner functions for the stabilized Fock states at a duration of approximately 50~$\mu$s, giving a state fidelity of 0.93, 0.86, and 0.77 for Fock states $|1\rangle$, $|2\rangle$, and $|3\rangle$, respectively. These measured fidelities are comparable with the estimated ones (0.96, 0.91 and 0.87) from analytical calculations in Supplementary Material~\cite{supplement}. The generated Fock states are preserved for up to approximately 10~ms, which is mainly limited by the classical control electronics used in the experiment and can potentially be extended indefinitely, indicating the successful and robust generation of the target multi-photon Fock states in the cavity. The quantum dynamics of the stabilization process can be fully described by a rate equation of the photon number populations during the evolution~\cite{supplement}, confirming the effectiveness of the non-unitary CSPA operations in adding photons to the cavity. The main infidelity of the stabilized Fock states arises from the mismatch of the coupling rates and decay rates (see Supplementary Material~\cite{supplement} for more details of the error analysis).

\begin{figure*}[htb]
    \includegraphics{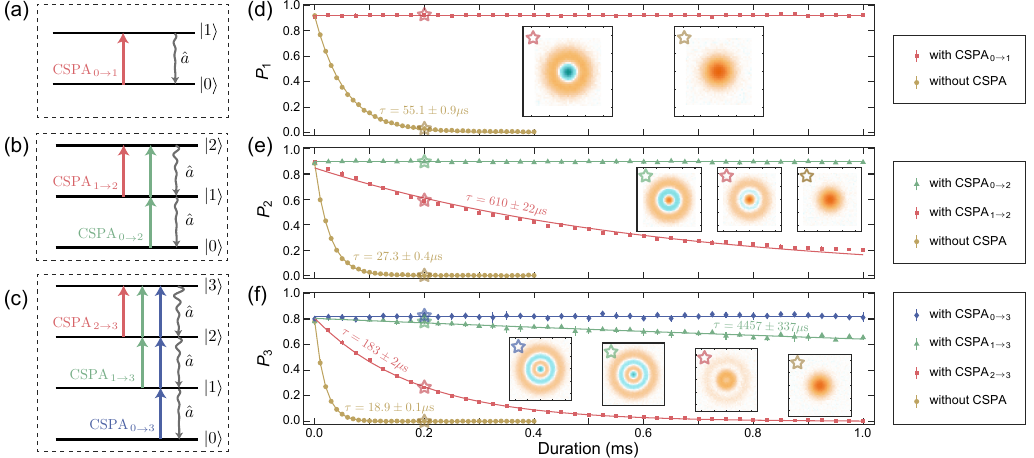}
    \caption{Fock state stabilization against multi-photon losses.
    (a)-(c) Energy level diagrams for illustrating the protection of Fock state $|1\rangle$ (a), $|2\rangle$ (b), and $|3\rangle$ (c) against losses of different number of photons.
   (d)-(f) Measured photon number populations (symbols) $P_1$ (d), $P_2$ (e), and $P_3$ (f) as a function of the stabilization duration when applying different sets of CSPA operations on corresponding initial Fock states. Error bars are standard deviations from repeated experiments. The decay curves are fitted to an exponential function $y = Ae^{-t/\tau}$ to extract the decay time constant $\tau$. Solid lines for the cases with $\mathrm{CSPA}_{0\rightarrow N}$ operation represent the average photon number population for the corresponding stabilized Fock states. The insets in (d)-(f) show the measured Wigner functions (with same axis scales and colorbars as that in Fig.~\ref{fig2}) at a duration of about 200~$\mu$s for each case. }
    \label{fig3}
\end{figure*}

To further validate the stabilization against multi-photon losses, we conditionally apply different configurations of CSPA operations after generating the initial Fock state $|N\rangle$ with $N=1,2,3$, as shown in Figs.~\ref{fig3}(a)-\ref{fig3}(c). We then measure the photon number population $P_N$ of the Fock state $|N\rangle$ as a function of the evolution time, as well as the measured Wigner snapshots at an evolution time of about $200~\mu$s for all cases, with the results shown in Figs.~\ref{fig3}(d)-\ref{fig3}(f). For example, in the case of initial Fock state $|3\rangle$, we apply non-unitary operations $\mathrm{CSPA}_{0\rightarrow 3}$, $\mathrm{CSPA}_{1\rightarrow 3}$, and $\mathrm{CSPA}_{2\rightarrow 3}$ to protect the cavity Fock state $|3\rangle$ against as large as 3-, 2-, and 1-photon losses, respectively. The measured photon number populations $P_3$ of the Fock state $|3\rangle$ for these three cases, as well as that without any CSPA operation, are shown in Fig.~\ref{fig3}(f) and fitted to an exponential function to extract the decay time constant. The Fock state shows slow decay rates due to the protection against additional photon losses when applying these CSPA operations. As a result, the measured decay time constants 55.1~$\mu$s, (27.3, 610)~$\mu$s, and (18.9, 183, 4457)~$\mu$s are consistent with the estimated ones 51~$\mu$s, (26, 639)~$\mu$s, and (17, 228, 4862)~$\mu$s for Fock states $|1\rangle$, $|2\rangle$, and $|3\rangle$, respectively~\cite{supplement}. These results demonstrate that the multi-photon Fock states in the cavity can indeed be preserved and protected from both single- and multi-photon-loss errors by the dissipation engineering method, highlighting the significant differences from previous works that only corrected single-photon-loss errors ~\cite{holland2015, gertler2021}.

\begin{figure}
    \includegraphics{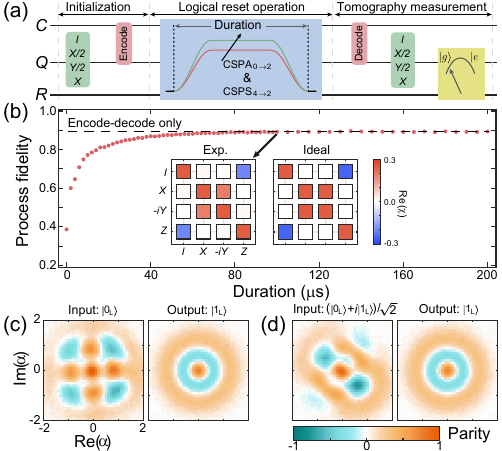}
    \caption{Logical reset operation of a binomially-encoded logical qubit. 
	(a) Experimental sequence to perform quantum process tomography to characterize the logical reset operation.
	(b) Measured process fidelity defined as $F=\left(\mathrm{Tr}\sqrt{\sqrt{\chi_\mathrm{ideal}}\chi_\mathrm{meas}\sqrt{\chi_\mathrm{ideal}} }\right)^2$, as a function of the duration of the reset operation. Here, $\chi_\mathrm{meas}$ and $\chi_\mathrm{ideal}$ are the measured and ideal process matrices for the logical reset operation, whose real parts are shown in the inset. Black dashed line represents the process fidelity with only the encoding and decoding processes.
	(c)-(d) Measured Wigner functions of the cavity states before and after the logical reset operation, which irreversibly transform the initial binomially-encoded logical states $|0\rangle_\mathrm{L}$ (c) and $(|0\rangle_\mathrm{L} + i|1\rangle_\mathrm{L})/\sqrt{2}$ (d) to a final state of $|1\rangle_\mathrm{L} $. }
    \label{fig4}
\end{figure}

Furthermore, our dissipation engineering method can also be extended to design a cascaded selective photon-subtraction (CSPS) operation, employing the third energy level of the transmon qubit~\cite{supplement}. This non-unitary CSPS operation can be utilized to protect the photonic states against photon-gain errors induced by thermal excitation in the oscillator. By combining the non-unitary CSPS operation with the CSPA operation, we can achieve greater flexibility in realizing arbitrary non-unitary quantum control over the oscillator.

To demonstrate the capabilities of these non-unitary operations, we present an example of their use for resetting a binomially-encoded logical qubit in cavity $C$. Here, the logical codewords are expressed as \{$|0\rangle_\mathrm{L} = (|0\rangle + |4\rangle)/\sqrt{2}$,  $|1\rangle_\mathrm{L} = |2\rangle$\}~\cite{michael2016}. The logical reset operation is achieved by implementing the non-unitary operations $\mathrm{CSPA}_{0\rightarrow 2}$ and $\mathrm{CSPS}_{4\rightarrow 2}$ simultaneously, which will irreversibly transform arbitrary logical states to the logical $|1\rangle_\mathrm{L}$ state. The process is characterized by the experimental sequence displayed in Fig.~\ref{fig4}(a). Note that due to the large decay rate of the Fock state $|4\rangle$ and the weak stabilization rate to add photons, it is unnecessary to apply the CSPS operation in our experiment. The measured process fidelity of the logical reset operation with the encoding and decoding processes, as a function of the total duration of the reset operation, is shown in Fig.~\ref{fig4}(b). As the duration exceeds 50~$\mu$s, the process fidelity approaches to a steady value of approximately 0.89, which is close to the process fidelity with only the encoding and decoding processes (0.894), indicating a high fidelity for the logical reset operation. Additionally, the measured process matrix at a duration of about 100~$\mu$s is presented in the inset of Fig.~\ref{fig4}(b), alongside the ideal matrix. Further verification of the reset operation is provided through measurements of the Wigner functions of the cavity states before and after the reset operation, with results shown in Figs.~\ref{fig4}(c)-\ref{fig4}(d).

In conclusion, we have demonstrated a dissipation engineering method to stabilize multi-photon Fock states in a superconducting microwave cavity, employing a cascaded selective photon-addition operation. This method represents a crucial development in protecting the cavity states against multi-photon losses beyond single-photon loss, which is critical for robust and high-precision quantum control. The experimental results confirm that the stabilized multi-photon Fock states can be preserved for a time much longer than that without any CSPA operations, opening up the possibility in quantum-enhanced metrology~\cite{deng2023} and dark matter search~\cite{Agrawal2023}. 
Additionally, our dissipation engineering method has been successfully employed for implementing a non-unitary operation for resetting a binomially-encoded logical qubit, indicating practical implications in error-correctable non-unitary operations of logical qubits.
The demonstrated dissipation engineering method can also be directly exploited to stabilize higher photon number states with improved device performance~\cite{Milul2023} and be applicable for optical photons and mechanical and acoustic wave phonons~\cite{wollack2022, vonlupke2022}, promising potential applications in quantum information processing with various bosonic modes.

\begin{acknowledgments}
We would like to thank Chang-Ling Zou and Fei Yan for helpful discussions. This work was supported by the Key-Area Research and Development Program of Guangdong Province (Grant No. 2018B030326001), the National Natural Science Foundation of China (Grant No. 12274198), the Shenzhen Science and Technology Program (Grant No. RCYX20210706092103021), the Guangdong Basic and Applied Basic Research Foundation (Grants No. 2024B1515020013, No. 2022A1515010324), the Guangdong Provincial Key Laboratory (Grant No. 2019B121203002), the Shenzhen-Hong Kong cooperation zone for technology and innovation (Contract NO. HZQB-KCZYB-2020050), and the Innovation Program for Quantum Science and Technology (Grant No. ZD0301703).
\end{acknowledgments}

\textit{Note added.}-- While we were preparing our manuscript, we noticed a similar implementation of selective photon addition operation, but lacking dissipation engineering~\cite{Kudra2022a}.

\end{document}


\title{Supplementary Material for ``Autonomous stabilization of Fock states in an oscillator against multiphoton losses"}

\author{Sai Li}
\thanks{These authors contributed equally to this work.}
\affiliation{Shenzhen Institute for Quantum Science and Engineering, Southern University of Science and Technology, Shenzhen 518055, China}
\affiliation{International Quantum Academy, Shenzhen 518048, China}
\affiliation{Guangdong Provincial Key Laboratory of Quantum Science and Engineering, Southern University of Science and Technology, Shenzhen 518055, China}

\author{Zhongchu Ni}
\thanks{These authors contributed equally to this work.}
\affiliation{Shenzhen Institute for Quantum Science and Engineering, Southern University of Science and Technology, Shenzhen 518055, China}
\affiliation{International Quantum Academy, Shenzhen 518048, China}
\affiliation{Guangdong Provincial Key Laboratory of Quantum Science and Engineering, Southern University of Science and Technology, Shenzhen 518055, China}
\affiliation{Department of Physics, Southern University of Science and Technology, Shenzhen 518055, China}

\author{Libo Zhang}
\author{Yanyan Cai}
\author{Jiasheng Mai}
\author{Shengcheng Wen}

\author{Pan Zheng}
\author{Xiaowei Deng}
\affiliation{Shenzhen Institute for Quantum Science and Engineering, Southern University of Science and Technology, Shenzhen 518055, China}
\affiliation{International Quantum Academy, Shenzhen 518048, China}
\affiliation{Guangdong Provincial Key Laboratory of Quantum Science and Engineering, Southern University of Science and Technology, Shenzhen 518055, China}

\author{Song Liu}
\email{lius3@sustech.edu.cn}

\author{Yuan Xu}
\email{xuyuan@iqasz.cn}
\affiliation{Shenzhen Institute for Quantum Science and Engineering, Southern University of Science and Technology, Shenzhen 518055, China}
\affiliation{International Quantum Academy, Shenzhen 518048, China}
\affiliation{Guangdong Provincial Key Laboratory of Quantum Science and Engineering, Southern University of Science and Technology, Shenzhen 518055, China}
\affiliation{Shenzhen Branch, Hefei National Laboratory, Shenzhen 518048, China}

\author{Dapeng Yu}
\affiliation{Shenzhen Institute for Quantum Science and Engineering, Southern University of Science and Technology, Shenzhen 518055, China}
\affiliation{International Quantum Academy, Shenzhen 518048, China}
\affiliation{Guangdong Provincial Key Laboratory of Quantum Science and Engineering, Southern University of Science and Technology, Shenzhen 518055, China}
\affiliation{Department of Physics, Southern University of Science and Technology, Shenzhen 518055, China}
\affiliation{Shenzhen Branch, Hefei National Laboratory, Shenzhen 518048, China}

 \maketitle

\section{Experimental device} 

The experiments conducted in this study utilize a circuit quantum electrodynamics~\cite{blais2021} device, similar to that in previous studies~\cite{Xu3D2020,Xu3D2021,ni2022,ni2023}. The device consists of a three-dimensional coaxial stub cavity~\cite{reagor2016} that dispersively couples to two superconducting transmon qubits~\cite{koch2007}, with each one coupling to a $\lambda/2$ stripline readout resonator~\cite{axline2016}. In our experiment, only the left qubit ($Q$) and its associated readout resonator ($R$) are used to implement quantum reservoir control for stabilizing multi-photon Fock states in cavity $C$. A schematic diagram of the device is shown in Fig.~1(b) in the main text. The device was placed inside a dilution refrigerator and cooled down to a base temperature of approximately 10~mK. The entire system is controlled using a set of commercial electronic equipment, primarily consisting of a Zurich Instruments high-density arbitrary waveform generator (HDAWG) and an ultra-high-frequency quantum analyzer (UHFQA). The wiring setup is similar to that in Ref.~\cite{Xu3D2021,ni2023}.

The system Hamiltonian can be described as
 \begin{eqnarray}
\label{H}
H&=&\omega_c \hat{a}^\dagger \hat{a} + \omega_q |e\rangle\langle e| + \omega_r \hat{r}^\dagger \hat{r} \notag\\
&-&\chi_{qc}|e\rangle\langle e|\hat{a}^\dagger \hat{a}-\chi_{qr}|e\rangle\langle e|\hat{r}^\dagger \hat{r}-\frac{\zeta_c}{2}\hat{a}^\dagger\hat{a}^\dagger \hat{a} \hat{a},
\end{eqnarray}
where $\omega_c$, $\omega_q$ and $\omega_r$ are the frequencies of the cavity, qubit and readout resonator, respectively. $\hat{a}$ and $\hat{r}$ are the annihilation operators of the cavity and readout resonator. $\chi_{qc}$ and $\chi_{qr}$ are the dispersive shifts between the qubit and the cavity and readout resonator, respectively. $\zeta_c$ is the self Kerr coefficient of the cavity. The main parameters and the coherence properties of the system are listed in Table~\ref{TableS1}.

\begin{table}[htb]
\caption{Device parameters and coherence properties.}
\begin{tabular}{p{7cm}<{\centering}p{1cm}<{\centering}}  
  \hline
  {Parameters} & {Values} \\
  \hline
  &\\[-0.9em]
  Cavity frequency $\omega_{c}/2\pi$~(GHz) & 6.366    \\
  &\\[-0.9em]
  Qubit frequency $\omega_{q}/2\pi$~(GHz) & 5.682    \\
  &\\[-0.9em]
  Readout frequency $\omega_{r}/2\pi$~(GHz) & 8.602 \\
  &\\[-0.9em]
  Cavity anharmonicity $\zeta_{c}/2\pi$~(kHz) & 150    \\
  &\\[-0.9em]
   $Q$-$C$ dispersive shift  $\chi_{qc}/2\pi$~(MHz) & 7.72    \\
  &\\[-0.9em]
   $Q$-$R$ dispersive shift $\chi_{qr}/2\pi$~(MHz) & 2.4  \\
  &\\[-0.9em]
   Qubit $T_1$~($\mu $s) & $18.6$    \\
  &\\[-0.9em]
   Qubit $T^*_2$~($\mu $s) & $25.6$     \\
  &\\[-0.9em]
   Qubit $T^\mathrm{echo}_2$~($\mu $s)  & $28.2$     \\
  &\\[-0.9em]
   Cavity $T_1$~($\mu $s)  & $51.9$     \\
  &\\[-0.9em]
   Cavity $T^*_2$~($\mu $s)  & $88.3$     \\
  &\\[-0.9em]
   Readout resonator decay rate $\kappa_r/2\pi$~(MHz) & $2.4$     \\
  &\\[-0.9em]
  \hline
\end{tabular}\vspace{-6pt}
\label{TableS1}
\end{table}


\section{Cascaded selective photon additions}

In our experiment, we implement the cascaded selective photon addition (CSPA) operations to autonomously stabilize Fock states against multi-photon losses in a harmonic oscillator. Here, the multi-photon losses include both the sequential loss of a single photon multiple times and the simultaneous loss of multiple photons at the same time through a non-linear process. Our engineered CSPA operations can effectively address both types of multi-photon-loss errors by sequentially adding photons one by one through continuous microwave drives. This is achieved by applying two sets of continuous-wave frequency combs to drive the system into a steady Fock state in the cavity, assisted by the fast dissipation of the readout resonator.

The first set of drives is resonant on the qubit frequencies when the cavity has $i$ photons, and realizes an oscillation between $|i,g,0\rangle$ and $|i,e,0\rangle$ states, which can be described by the drive Hamiltonian
\begin{eqnarray}
\label{HQd}
H_{d1}&=&\Omega_i|i,g,0\rangle\langle i, e,0| + \textrm{H.c.},
\end{eqnarray}
where $\Omega_i$ represents the Rabi rate of the qubit drive.

The second set of drives targets the transition $|i,e,0\rangle\leftrightarrow|i+1,g,1\rangle$ through a four-wave mixing process~\cite{Gertler2021}. This process can be described by driving a Josephson junction, whose non-liner Hamiltonian is expressed as $H_\mathrm{nl} = -E_J \left( \cos{\hat{\varphi}} + \frac{\hat{\varphi}^2}{2} \right)$, where $E_J$ represents the Josephson junction energy, and $\hat{\varphi} = \phi_c\hat{a} + \phi_q\hat{q} + \phi_r\hat{r}+ \textrm{H.c.}$ is the phase operator of the Josephson junction. $\hat{q}$ is the annihilation operator of the qubit mode and $\phi_c$, $\phi_q$ and $\phi_r$ are the zero-point phase fluctuations across the junction of the cavity, qubit and readout resonator modes, respectively. To account for the four-wave mixing pumps with drive frequencies $\omega_i$ and drive amplitudes $\epsilon_i$, a displacement unitary transformation $ \hat{q}\rightarrow \hat{q} + \epsilon_i e^{i\omega_i t}$ is performed to achieve a drive Hamiltonian
\begin{eqnarray}
\label{HFd}
H_{d2}&=&-E_J\phi^2_q\phi_c\phi_r \epsilon_i\hat{q}\hat{a}^\dagger\hat{r}^\dagger+ \mathrm{H.c.},
\end{eqnarray}
when the frequency matching condition $\omega_i = \omega_c + \omega_r - \omega_q$ is satisfied. In addition, the four-wave mixing pump will also induce a Stark shift of the drive frequency, which has been carefully calibrated and mitigated in the experiment.

\begin{figure}[t]
	\centering
	\includegraphics{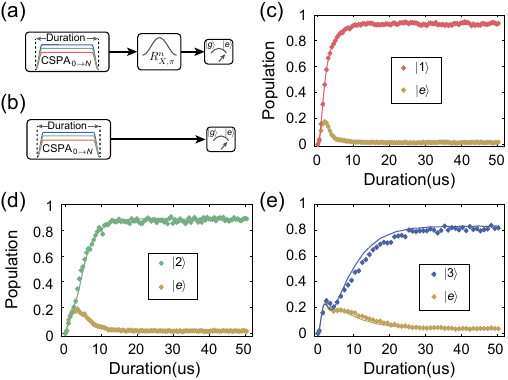}
	\caption{ Calibrating the coupling rates of the CSPA operations. (a) and (b) Experimental sequence for detecting the photon number population and the qubit excitation probability during the $\mathrm{CSPA}_{0\rightarrow N}$ process. (c-e) Qubit excitation populations and cavity photon number populations for the $\mathrm{CSPA}_{0\rightarrow N}$ operation to stabilize Fock state $|1\rangle$ (c), $|2\rangle$ (d), and $|3\rangle$ (e), respectively. The experimental results are fitted with numerical simulations to extract the coupling rates of the two-sets of drives. }
	\label{FigCalrate}
\end{figure}

Combining the two sets of continuous-wave drives, we can achieve a total drive Hamiltonian
\begin{eqnarray}
\label{Hmd}
H_d&=&\Omega_i |i,g,0\rangle\langle i,e,0|\notag\\
&+&J_i |i,e,0\rangle\langle i+1,g,1|+\text{H.c.},
\end{eqnarray}
where $J_i = -E_J\phi^2_q\phi_c\phi_r\sqrt{i+1}\epsilon_i$ represents the effective coupling rate between the states $|i,e,0\rangle$ and $|i+1,g,1\rangle$. 

Combining with the fast spontaneous decay of the reservoir resonator which irreversibly transfers the state $|i+1,g,1\rangle$ to $|i+1,g,0\rangle$, the system evolves to a steady Fock state $|i\rangle$ in the cavity, realizing a selective photon-addition operation $\Lambda_{i\rightarrow i+1} = |i+1\rangle \langle i|$ between two adjacent energy levels in the cavity, given that $\kappa_c < \Omega_i < J_i < \kappa_r <\chi_{qc}$. It is important to note that choosing a larger driving strength would increase the stabilization rate, but at the expense of reducing the selectivity of the microwave drive, given a fixed dispersive shift $\chi_{qc}$, which in turn could limit the final stabilized fidelity. In our experiment, we choose these parameters based on numerical simulations to identify an optimal compromise that balances the trade-off between enhancing stabilization rates and maintaining selectivity, thereby maximizing the stabilized state fidelity.

Simultaneously applying all the drives of the two sets of continuous-frequency combs, we can achieve a total Hamiltonian of  
\begin{eqnarray} 
\label{Hd}
H_d &=& \sum_{i=m}^{n-1}{\left(\Omega_i |i,e,0\rangle \langle i,g,0| + J_i |i+1,g,1\rangle \langle i,e,0| \right)} \notag\\
&+& \mathrm{H.c.},
\end{eqnarray}
in a tailored photon number subspace. By combining with the spontaneous decay of the reservoir resonator, we can achieve the non-unitary operation $\mathrm{CSPA}_{m\rightarrow n} = \sum_{i=m}^{n-1}{|i+1\rangle \langle i |}$, which is similar to the conjugate operator of $\hat{a}$ in a truncted photon number subspace. The only difference is the absence of the photon-number-dependent coefficient, which has little effects for the Fock state stabilization. Continuously implementing this operation $n-m$ times $(\Pi_{j=1}^{n-m}{\mathrm{CSPA}_{m\rightarrow n}} = |n\rangle\langle m|)$ would autonomously stabilize the Fock state $|n\rangle$ against $n-m$ photons in the cavity. It is noteworthy that the CSPA operation can be viewed as a generalization of the previous parity recovery operation, where the photon-addition operators are only applied on odd photon numbers to correct single-photon-loss error~\cite{Gertler2021}.

In order to calibrate the coupling rates of the drive Hamiltonian, we perform a $\mathrm{CSPA}_{0\rightarrow N}$ operation to stabilize a Fock state $|N\rangle$ from an initial vacuum state. The photon number population and qubit excited state population are measured during the evolution of the CSPA operation, with the experimental sequence shown in Figs.~\ref{FigCalrate}(a-b). The experimental results are shown in Figs.~\ref{FigCalrate}(c-e) for Fock state $|N\rangle$ with $N=1,2,3$. 

The coupling rates are obtained by performing numerical simulations based on the drive Hamiltonian in Eq.~(\ref{Hd}) to fit these results. However, in practice, the multi-component continuous-wave drives will introduce spurious couplings into the drive Hamiltonian~\cite{Gertler2021}, which can be expressed as
\begin{eqnarray}
\label{HFulld}
H'_d&=&\sum^{N-1}_{j=0}e^{[\text{i} (j\chi_{qc})t]}\sum^{N}_{i=0}(\Omega_j |i,g,0\rangle\langle i,e,0|\notag\\
&+&J_j \sqrt{\frac{i+1}{j+1}}|i,e,0\rangle\langle i+1,g,1|+\text{H.c.}).
\end{eqnarray}
This equation includes both the resonant and off-resonant dynamics of each photon-number state $(i = 0$ to $N)$ under the effect of all tones $(j = 0$ to $N-1)$ in the CSPA operation. In our experiment, we use this modified Hamiltonian to perform numerical simulations to fit the experimental data in Fig.~\ref{FigCalrate}. The obtained coupling rates $\Omega_i$ and $J_i$ for the two sets of continuous-wave frequency comb drives are summarized and listed in Table.~\ref{TableS1}. These parameters are used to perform numerical simulations of the experiments in Fig.~2 in the main text.

Based on these coupling rates, we could obtain the stabilizing rate $\lambda_i$ for adding a photon to state $|i\rangle$, which can be calculated by $\lambda_i = 1/(1/\Omega_i+1/J_i+1/\kappa_r)$~\cite{holland2015}. As a result, the stabilizing rates are $\lambda_0/2\pi = 68.7$~kHz for Fock state $|1\rangle$, $(\lambda_0, \lambda_1)/2\pi = (68.7, 68.1)$~kHz for Fock state $|2\rangle$, and $(\lambda_0, \lambda_1, \lambda_2)/2\pi = (66.3, 66.7, 67.9)$~kHz for Fock state $|3\rangle$, respectively.

\begin{table}[b]
\caption{Coupling rates of the two sets of drives of the $\mathrm{CSPA}_{0\rightarrow N}$ operation.}
\begin{tabular}{p{4.4 cm}<{\centering}|p{2cm}<{\centering}p{2cm}<{\centering}p{2cm}<{\centering}}  
  \hline
&\\[-0.9em]
   \diagbox{CSPA operation}{Coupling rates} & $\Omega/2\pi$ (kHz) & $J/2\pi$ (kHz)\\
&\\[-0.9em]
  \hline
  &\\[-0.9em]
  $\textrm{CSPA}_{0\rightarrow 1}$ & $\Omega_0 = 86$   & $J_0 = 400$ \\
  \hline
  &\\[-0.9em]
  \multirow{2}{*}{$\textrm{CSPA}_{0\rightarrow 2}$} & $\Omega_0 = 86$   & $J_0 = 400$ \\
                                   & $\Omega_1 = 86$   & $J_1 = 380$ \\
\hline
  &\\[-0.9em]
   \multirow{3}{*}{$\textrm{CSPA}_{0\rightarrow 3}$} & $\Omega_0 = 86$  & $J_0 = 330$ \\
                                    & $\Omega_1 = 86$   & $J_1 = 341$ \\
                                    & $\Omega_2 = 87$   & $J_2 = 356$ \\
  \hline
\end{tabular}\vspace{-6pt}
\label{TableS2}
\end{table}

\section{Rate equation for describing the dynamics of the CSPA operation}
The quantum dynamics of the Fock state stabilization process can be described by a rate equation of the photon number populations. The time evolution of the photon number population ${P_i}(t)$ of Fock state $|i\rangle$ during CSPA operation follows the equation: 
\begin{eqnarray}
\label{rateF}
\frac{d{P_i}(t)}{dt} &=& \lambda_{i-1}{P_{i-1}}(t)-\lambda_{i}{P_{i}}(t) \notag\\
&-&i\kappa_c{P_{i}}(t)  +  (i+1)\kappa_c{P_{i+1}}(t).
\end{eqnarray}
where $\kappa_c/2\pi = 3.1$ kHz is the cavity decay rate. This rate equation describes the time evolution of the photon number populations and includes the photon addition and spontaneous decay processes in the cavity. Alternatively, this rate equation can be written in a matrix form as ${d\vec{P}(t)}/{dt} = \Gamma\cdot\vec{P}(t)$, where $\Gamma$ is the rate matrix that can be experimentally calibrated. 


The differential equation of the photon number population evolution can be solved under specific initial conditions to obtain the photon number populations ${P_i}(t)$. This allows us to determine the steady Fock state population ${P_N}$ for the $\mathrm{CSPA}_{0\rightarrow N}$ operation, as shown in Fig.~(2) in the main text. Additionally, we can use this rate equation to describe the protection of the Fock state $|N\rangle$ against multi-photon losses in the cavity, as shown in Fig.~(3) in the main text.

When the photon number populations reach stability with the $\textrm{CSPA}_{0\rightarrow N}$ process, We have ${d\vec{P}(t)}/{dt} = \Gamma\cdot\vec{P}(t) = 0$. From this condition, we can derive the recursive relations as
\begin{eqnarray}
\label{PR}
P_{i} = (i+1)\kappa_c/\lambda_i\cdot P_{i+1}.
\end{eqnarray}
Using these recursive relations, we can determine the fidelity of the stabilized photon number state $|N\rangle$, given by
\begin{eqnarray}
\label{PN}
F_N = \left(1 + \sum^{N-1}_{i=0}\prod^N_{m = N-i}\frac{m\kappa_c}{\lambda_{m-1}}\right)^{-1}.
\end{eqnarray}
From this equation, the fidelities of the final steady states of the $\mathrm{CSPA}_{0\rightarrow N}$ are calculated as $F_1  = 0.957$, $F_2 = 0.913$, and $F_3 = 0.869$ for stabilized Fock states $|1\rangle$, $|2\rangle$, and $|3\rangle$, respectively. These results are consistent with the numerical simulations, confirming the effectiveness of the rate equation description of the $\mathrm{CSPA}_{0\rightarrow N}$ operation.

\begin{table}[t]
\caption{Estimated decay time $\tau$ for $|N\rangle$ with CSPA operations.}
\setlength{\tabcolsep}{12pt}
\renewcommand{\arraystretch}{1.0}
\begin{tabular}{@{}c|ccc}
  \hline
   \diagbox[width = 13em,trim=l]{CSPA operations}{ $\tau$ (us)}{Initial state} & $|1\rangle$ & $|2\rangle$ & $|3\rangle$\\
  \hline
  &\\[-0.9em]
   Without CSPA & 51 & 26 & 17 \\
    \hline
  &\\[-0.9em]
    $\mathrm{CSPA}_{N-1\rightarrow N}$  & - & 639 & 228 \\
   \hline
  $\mathrm{CSPA}_{N-2\rightarrow N}$  & - & - & 4862\\
  \hline
\end{tabular}\vspace{-6pt}
\label{TableS3}
\end{table}

To analyze the protection of the Fock state $|N\rangle$ against different photon number losses with the $\mathrm{CSPA}_{i\rightarrow N}$ operation, we solve the rate equation ${d\vec{P}(t)}/{dt} = \Gamma\cdot\vec{P}(t)$ using the following approach. First, we assume a set of general solutions for the photon number population as
\begin{eqnarray}
\label{solutions}
\vec{P}(t) = e^{\eta t}\vec{r},
\label{s13}
\end{eqnarray}
where $\eta$ and $\vec{r}$ are to be determined. Next, we substitute this solution into the rate equation and obtain: 
\begin{eqnarray}
\label{det1}
(\Gamma - \eta I )\vec{r} = 0.
\end{eqnarray}
This equation should satisfy the condition $\text{det}(\Gamma - \eta I )= 0$. This condition allows for determining the characteristic values $\eta_i$, and  derive the characteristic vectors $\vec{r}_i$, which is actually the diagonalization of the rate matrix. Finally, the photon-number population can be expressed as
\begin{eqnarray}
\label{PN2}
\vec{P}(t) = \sum_i c_i \vec{r}_i e^{\eta_i t},
\end{eqnarray}
where $c_i$ is determined by the experimental system. In our experiment, for the initial state $|N\rangle$, we have $P_{i}\approx 0~(i<N)$ since the effective pumping rate $\lambda_i$ is much larger than the decay rate $(i+1)\kappa_c$. Thus, we can deduce that the decay rate of $P_{N}$ is determined by $\eta_N$. As a result, we can estimate the decay time constant $\tau$ for protecting Fock $|N\rangle $ against different number of photon losses. The predicted $\tau$ are listed in Table ~\ref{TableS3}, which have good agreements with the experimental results shown in Fig.~3 in the main text, validating the effectiveness of the rate equation to describe the CSPA operation.

\section{Error analysis of the stabilized Fock states}

In this section, we analyze the main error sources and their contributions to the infidelities of the stabilized Fock states. The dominant error sources are identified through numerical simulations and analyzed as follows:

1. The ratio of the effective pumping rate $\lambda_i$ and the cavity decay rate $\kappa_c$ sets the innate upper limit on the fidelity of the stabilized Fock states~\cite{holland2015}. 

2. The spurious couplings of the multi-component drives induce undesirable transitions during the CSPA operation. These spurious effects have been taken into account in numerical simulations by utilizing the modified drive Hamiltonian in Eq.~(\ref{HFulld}). 

3. Cavity heating events can occur due to the qubit thermal excitation and the presence of spurious drives. For instance, in the case for stabilizing Fock state $|1g0\rangle$, the system can be heated to $|1e0\rangle$ due to the qubit thermal excitation with a rate of about 0.06 kHz. The continuous-wave mixing pulse then drives $|1e0\rangle$ to $|2g1\rangle$ off-resonantly with a detuning of $\chi_{qc}$. And finally the system evolves to $|2g0\rangle$ because of the fast spontaneous decay of the reservoir resonator, thus resulting in cavity heating effect.

4. The imperfect compensation of the residual Stark shifts can result in infidelities of the final steady Fock states during the CSPA operations. These residual Stark shifts are also scrutinized in numerical simulations.

\begin{table}[b]
\caption{Numerically estimated errors for stabilizing Fock $|N\rangle$.}
\setlength{\tabcolsep}{11pt}
\renewcommand{\arraystretch}{1.0}
\begin{tabular}{@{}c|ccc}
  \hline
&\\[-0.9em]
   \diagbox[width = 13em,trim=l]{Error sources}{Target state} & $|1\rangle$ & $|2\rangle$ & $|3\rangle$\\
&\\[-0.9em]
  \hline
  &\\[-0.9em]
  $\text{ Pumping rate $\lambda_{i}$ limitation}$ & 0.043   & 0.083 & 0.118\\ 
  &\\[-0.9em]
  $\text{Spurious drive}$ & 0.010   & 0.020 & 0.024\\
  &\\[-0.9em]
   $\text{Cavity heating}$ & 0.007 & 0.004 & 0.002 \\
  &\\[-0.9em]
  $\text{Residual Stark shift}$ & 0.012   & 0.003 & 0.016\\
  &\\[-0.9em]
  $\text{Cavity decay in detection}$ & 0.003 & 0.006 & 0.009\\
\hline
  &\\[-0.9em]
   $\text{Total errors }$ & 0.075 & 0.117 &  0.170\\
  \hline
\end{tabular}\vspace{-6pt}
\label{TableS4}
\end{table}

5. The detection errors for measuring the photon number populations come from the cavity decay during the long selective $\pi$ pulse. 

\begin{figure}[t]
	\centering
	\includegraphics{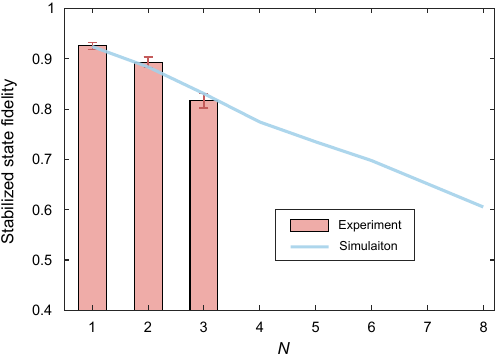}
	\caption{The state fidelity of the stabilized Fock states with various photon number $N$ from numerical simulations (solid line) and experiment (bar). }
	\label{FigLN}
\end{figure}

Finally, We estimate the contributions of all these error sources through numerical simulations, and the results are listed in Table ~\ref{TableS4}. Additionally, we plot the state fidelity of the stabilized Fock states with various photon numbers from numerical simulations in Fig.~\ref{FigLN}.

 \begin{figure}[t]
	\centering
	\includegraphics{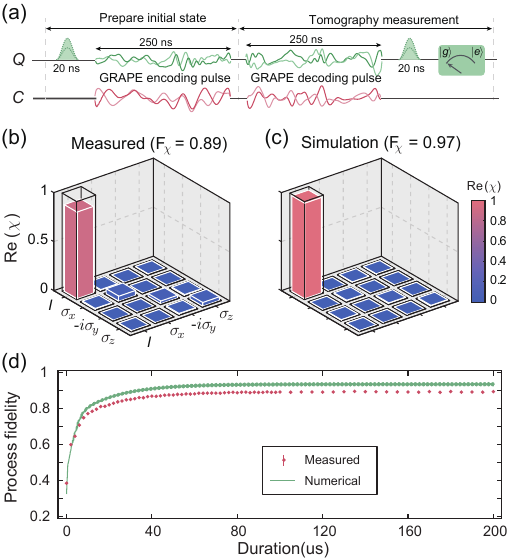}
	\caption{The calibration of the encoding and decoding processes for resetting a binomial logical qubit in the cavity. (a) Quantum process tomography experimental sequence for characterizing the encoding and decoding processes of a binomial logical qubit. (b) Experimentally measured and (c) numerically simulated process $\chi$ matrices of the encoding and decoding processes. (d) Measured and simulated process fidelities as a function of the duration of the CSPA operation for resetting a binomially-encoded logical qubit in the cavity. A small difference of about 0.04 for the steady values may come from the GRAPE pulse distortion of the decoding process.
}
	\label{FigReset}
\end{figure}

\section{The encoding and decoding processes of a binomial logical qubit}

In our experiment, we utilize the gradient ascent pulse engineering (GRAPE) technique~\cite{khaneja2005} to encode and decode the binomial logical qubit \{$|0\rangle_\mathrm{L} = (|0\rangle + |4\rangle)/\sqrt{2}$,  $|1\rangle_\mathrm{L} = |2\rangle$\}~\cite{Michael2016}. However, imperfections arise in the encoding and decoding processes due to pulse distortions and dissipation during the GRAPE operations~\cite{mizuno2023}. Quantum process tomography~\cite{nielsen2010} experiment is performed to calibrate the fidelity of only the encoding and decoding processes, with the pulse sequence shown in Fig.~{\ref{FigReset}}(a). The measured process matrix is presented in Fig.~{\ref{FigReset}}(b), exhibiting a process fidelity of 0.894. We also perform a numerical simulation of this process, giving a simulated process fidelity of 0.97 for the process matrix shown in Fig.~{\ref{FigReset}}(c). The fidelity of the measurement result deviates from that of the numerical simulation by about 0.08 due to the distortions of the GRAPE pulse, which is difficult to account for in the numerical simulation. Since the optimization procedures are the same for all GRAPE pulses, each one of the encoding and decoding processes has an additional error of about 0.04 comparing with the numerical simulations.

The CSPA operation for resetting the binomially-encoded logical qubit is robust to the imperfection of the encoding process. 
Even if the encoded state is imperfect, the CSPA operation can always stabilize the cavity state to the target logical state $|1\rangle_\text{L}$. However, the imperfection of the decoding process will result in errors for characterizing the logical reset operation. The numerically simulated process fidelity as a function of the duration of the CSPA operation for the logical reset operation is shown in Fig.~{\ref{FigReset}}(d), together with the measured process fidelity. A slight discrepancy of about 0.04 between them for the steady values may result from the additional errors of the decoding process due to the GRAPE pulse distortion. 

In our experiment, the logical reset operation achieves a steady value when the operation duration exceeds 50 $\mu$s. The primary limitation of the logical reset operation may be the slow transition rates of the non-unitary CSPA and CSPS operations, which could be overcome by increasing the dispersive interaction $\chi_{qc}$ and the decay rate $\kappa_r$. It is worth mentioning that an alternative approach to implement the logical reset operation involves first dumping the cavity photons into the readout resonator through parametric conversion~\cite{NP2017Pfaff} and then generating Fock state $|2\rangle$ using the quantum optimal control technique~\cite{khaneja2005}. However, our demonstrated scheme offers potential advantages over this alternative proposal. Specifically, our demonstrated logical reset operation stabilizes the final Fock state without relying on the complex and non-scalable numerical optimal control and approaches a steady value with more robustness. Additionally, non-unitary logical reset operations can also be achieved by employing measurement and feed-forward strategies~\cite{bluvstein2024a}, which are distinct from the dissipation engineering method used here.

\begin{figure}[b]
	\centering
	\includegraphics{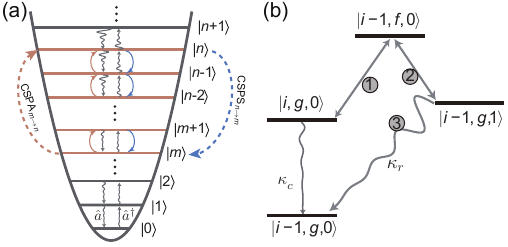}
	\caption{Schematic of the cascaded selective photon subtractions. (a) Illustration of arbitrary non-unitary quantum operations over the oscillator realized by the CSPA and CSPS operations in a truncated photon number subspace.
 (b) Energy level diagram of the $C$-$Q$-$R$ system for achieving the CSPS operation, which is realized by combining two sets of continuous-wave frequency drives (\ding{172} and \ding{173}) and the fast spontaneous decay of the readout resonator (\ding{174}). 
}
	\label{FigCSPS}
\end{figure}

\section{Cascaded selective photon subtraction}
To implement arbitrary non-unitary quantum operations on the oscillator within a truncated photon number subspace shown in Fig.~{\ref{FigCSPS}}(a), we also design a cascaded selective photon-subtraction (CSPS) operation using the third energy level state $|f\rangle$ of the transmon qubit. The implementation of the CSPS operation is illustrated in Fig.~{\ref{FigCSPS}}(b), which involves three steps: 1) achieving resonant oscillations between the state $|i,g,0\rangle$ and $|i-1,f,0\rangle$ by a storage-cavity-assisted Raman transition process~\cite{pechal2014, zeytinoglu2015}; 2) implementing resonant drives on the transition $|i-1,f,0\rangle \leftrightarrow |i-1,g,1\rangle$ using a similar Raman transition process between the qubit and the readout resonator; 3) utilizing the fast spontaneous decay of the reservoir resonator $R$ for irreversibly transferring the state $|i-1,g,1\rangle$ to $|i-1,g,0\rangle$. Finally, we can construct a photon-subtraction operation $\hat{\Lambda}_{i\rightarrow i-1} = |i-1\rangle \langle i |$ that subtracts a single photon from the Fock state $|i\rangle$ in the cavity. By simultaneously applying a set of these drives, the non-unitary CSPS operation $\mathrm{CSPS}_{n\rightarrow m}$ can be realized in a truncated photon number subspace. This non-unitary CSPS operation can be employed to protect the photon number states against photon-gain errors caused by thermal excitation in the oscillator. By combining the CSPS and CSPA operations, we achieve greater flexibility in arbitrary non-unitary control of harmonic oscillators, promising potential applications in quantum information processing.

\section{Mitigating the detection errors}
In our experiment, the stabilized cavity states are characterized by measuring the photon number populations and Wigner functions assisted by the superconducting qubit. In order to characterize the performance of the stabilization process, we calibrate and mitigate the detection errors during the measurement of the stabilized states.

\begin{figure}[t]
	\centering
	\includegraphics{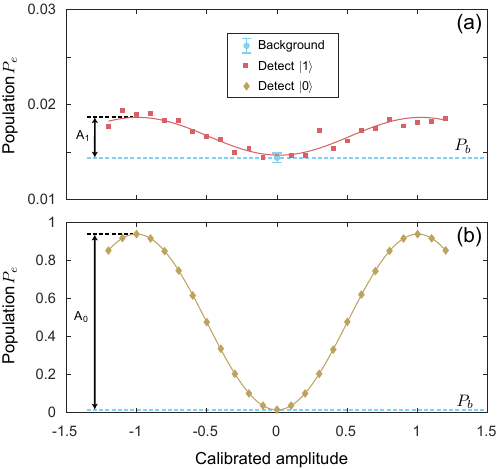}
	\caption{The calibration of detection errors for measuring the photon number populations of the cavity states. (a) Measured qubit excited state population as a function of the pulse amplitude of the selective Rabi drive with a detuning of 0. The background population of the qubit readout is measured without performing any pre-rotations. (b) Measured qubit excited state population as a function of the pulse amplitude of the selective Rabi drive with a detuning of $\chi_{qc}$. Fitting the two oscillations with a cosine function gives the oscillations amplitudes $A_0$ and $A_1$, respectively. 
}
	\label{FigDetection}
\end{figure}

We first calibrate the qubit readout errors by initializing the qubit in each computational basis state and measuring the assignment fidelities as $F_g = 0.985$ and $F_e=0.952$ for the qubit prepared in $|g\rangle$ and $|e\rangle$, respectively. To mitigate the qubit readout errors, we use a calibration matrix
 \begin{eqnarray}
\label{Mmatrix}
\mathcal{M} = \left(
\begin{array}{cc}
F_g & 1-F_e\\
1-F_g & F_e
\end{array}
\right).
\end{eqnarray}
to obtain the corrected qubit probability $\vec{P}_\mathrm{corr} = (P_g, P_e)^T$ by $\vec{P}_\mathrm{corr} = \mathcal{M}^{-1}\cdot\vec{P}_*$, where $\vec{P}_*$ represents the uncorrected measurement result.

For the measurement of the photon number populations of the cavity state, we perform a selective $\pi$ pulse with a variable detuning of $n\chi_{qc}$ before the qubit readout. The measured qubit excited state population corresponds to the photon number population $p_n$ of the cavity state. In order to mitigate the qubit decoherence errors during the selective $\pi$ pulse, we perform two selective Rabi oscillation experiments for the calibration~\cite{christopher2020}. For an initial vacuum state in the cavity, we first measure the qubit excited state probability as a function of the amplitude of the selective Rabi pulse applied on the qubit. With choosing a detuning of 0 and $\chi_{qc}$ for the selective Rabi drive, the measured Rabi oscillations are shown in Fig.~\ref{FigDetection}, with the oscillation amplitudes $A_0$ and $A_1$ proportional to the cavity photon number populations $p_0$ and $p_1$, respectively. The ratio of the two amplitudes gives a thermal population of about 0.5\% of the cavity.

Under the assumption that there is no photon number dependence to the decoherence rate of the qubit, we can extract the corrected photon number population $\vec{P}$ of the cavity state via
\begin{eqnarray}
\label{PE}
\vec{P} = \frac{\vec{P}_e- P_b}{A_0+A_1},
\end{eqnarray}
where $\vec{P}_e$ is the measured qubit excited state probabilities for selective $\pi$ rotations conditional on all the photon numbers in the cavity, and $P_b$ is the background population of the qubit readout. Note that the extracted photon number populations may be inaccurate during the intermediate evolution of the Fock state stabilization process, since the qubit is not always in the ground state during the evolution.

In order to calibrate the detection errors for measuring the Wigner functions of the stabilized cavity states, we implement a parity measurement of the initial vacuum state in the cavity. The parity measurement sequence is $X/2\rightarrow C_\pi \rightarrow \pm X/2$ with $C_\pi = \ket{g}\bra{g}\otimes I + \ket{e}\bra{e}\otimes e^{i\pi a^\dagger a}$ representing the qubit-state-dependent conditional phase gate of the cavity. The measured qubit ground state populations $P_\pm$ are utilized for obtaining the zero point Wigner function of the initial vacuum state as $W(0) = P_- - P_+ = 0.924$, which quantifies the readout errors during the Wigner tomography measurement. In order to mitigate the detection errors during the parity measurement of the stabilized Fock states in the cavity, we use this value to normalize the measured Wigner function $W(\alpha)$ via $W_\mathrm{corr}(\alpha) = W(\alpha)/W(0)$ based on the assumption that the parity measurements have little influence on finite photon numbers in the cavity.